\documentstyle [12pt,amsfonts] {article}
\input epsf

\newcommand{\beq}{\begin{equation}}
\newcommand{\eeq}{\end{equation}}
\newcommand{\beqs}{\begin{eqnarray}}
\newcommand{\eeqs}{\end{eqnarray}}

\catcode`@=11
\@addtoreset{equation}{section}
\@addtoreset{equation}{subsection}
\def\theequation{\ifnum\value{section}=0 \arabic{equation}\ignorespaces
\else \ifnum\value{section}=-1 A.\arabic{equation}\ignorespaces
\else \ifnum\value{subsection}=0 \thesection.\arabic{equation}\ignorespaces
\else \thesection.\arabic{subsection}.\arabic{equation}\ignorespaces
                           \fi
                      \fi
                 \fi}
\catcode`@=12

\begin{document}

\def\thefootnote{\fnsymbol{footnote}}

\baselineskip 6.0mm

\vspace{4mm}

\begin{center}

{\Large \bf $T=0$ Partition Functions for Potts Antiferromagnets on Square
Lattice Strips with (Twisted) Periodic Boundary Conditions} 

\vspace{8mm}

\vspace{8mm}

\setcounter{footnote}{0}
Norman Biggs\footnote{email: n.l.biggs@lse.ac.uk}

\vspace{4mm}

Centre for Discrete and Applicable Mathematics \\
London School of Economics \\
London  WC2A 2AE \\
UK \\

\vspace{4mm}

\setcounter{footnote}{6}
Robert Shrock\footnote{email: robert.shrock@sunysb.edu} \\
Institute for Theoretical Physics \\
State University of New York       \\
Stony Brook, N. Y. 11794-3840  \\
USA \\

\vspace{4mm}

{\bf Abstract}
\end{center}

We present exact calculations of the zero-temperature partition function
for the $q$-state Potts antiferromagnet (equivalently, the chromatic
polynomial) for two families of arbitrarily long strip graphs of the square
lattice with periodic boundary conditions in the transverse direction and (i)
periodic (ii) twisted periodic boundary conditions in the longitudinal
direction, so that the strip graphs are embedded on a (i) torus (ii) Klein
bottle.  In the limit of infinite length, we calculate the exponent of the 
entropy, $W(q)$, show it to be the same for (i) and (ii), and determine its 
analytic structure.  

\vspace{16mm}

\pagestyle{empty}
\newpage

\pagestyle{plain}
\pagenumbering{arabic}
\renewcommand{\thefootnote}{\arabic{footnote}}
\setcounter{footnote}{0}

The chromatic polynomial $P(G,q)$ counts the number of ways that one can color
a graph $G$ with $q$ colors such that no two adjacent vertices have the same
color \cite{birk} (for reviews, see \cite{rrev}-\cite{bbook}).  The least 
positive integer $q$ for which $P(G,q)$ is nonzero is the chromatic number, 
$\chi(G)$. 
Besides its intrinsic mathematical interest, the chromatic polynomial has an
important connection with statistical mechanics since it is the
zero-temperature partition function of the $q$-state Potts antiferromagnet (AF)
\cite{potts,wurev} on $G$: $P(G,q)=Z(G,q,T=0)_{PAF}$.  The Potts AF exhibits
nonzero ground-state entropy $S_0 \ne 0$ (without frustration) for sufficiently
large $q$ on a given lattice graph and is thus an exception to the third law of
thermodynamics. This is equivalent to a ground state
degeneracy per site $W > 1$, since $S_0 = k_B \ln W$.  Denoting the number of
vertices of $G$ as $n=v(G)$ and assuming a reasonable definition of 
$\{G\}=\lim_{n \to \infty}G$, we
have\footnote{\footnotesize{ At certain special points $q_s$ (typically
$q_s=0,1,.., \chi(G)$), one has the noncommutativity of limits $\lim_{q \to
q_s} \lim_{n \to \infty} P(G,q)^{1/n} \ne \lim_{n \to \infty} \lim_{q \to
q_s}P(G,q)^{1/n}$, and hence it is necessary to specify the order of the limits
in the definition of $W(\{G\},q_s)$ \cite{w}.  We use the first order of 
limits here; this has the advantage of removing certain isolated 
discontinuities in $W$.}} $W(\{G\},q)=\lim_{n \to \infty} P(G,q)^{1/n}$. 
Since $P(G,q)$ is a polynomial, one can generalize $q$ from ${\mathbb Z}_+$ to
${\mathbb C}$.  The zeros of $P(G,q)$ in the complex $q$ plane are called 
chromatic zeros; a subset of these may form an accumulation set in the $n \to 
\infty$ limit, denoted ${\cal B}$, which is the continuous locus of points 
where $W(\{G\},q)$ is nonanalytic \cite{bkw,read91,w} (${\cal B}$ may be 
null, and $W$ may also be nonanalytic at certain discrete points). 
The maximal region in the complex $q$ plane to which one
can analytically continue the function $W(\{G\},q)$ from physical values where
there is nonzero ground state entropy is denoted $R_1$.  The maximal value of
$q$ where ${\cal B}$ intersects the (positive) real axis is
labelled $q_c(\{G\})$. 

We consider strips of the square lattice with arbitrary length $L_x=m$ vertices
and fixed width $L_y$ vertices (with the longitudinal and transverse directions
taken to be $\hat x$ and $\hat y$). The chromatic polynomials for the cyclic
and M\"obius strip graphs of the square lattice were calculated for $L_y=2$ in
\cite{bds} (see also \cite{bm}-\cite{matmeth}) and for $L_y=3$ in
\cite{wcy,pm}. After studies of the chromatic zeros for $L_y=2$ in
\cite{bds,readcarib88,read91}, $W$ and ${\cal B}$ were determined for this case
in \cite{w} and for $L_y=3$ in \cite{wcy}.  An important question concerns the
effect of boundary conditions (BC's), and hence graph topology, on $P$, $W$,
and ${\cal B}$. We use the symbols FBC$_y$ and PBC$_y$ for free and periodic
transverse boundary conditions and FBC$_x$, PBC$_x$, and TPBC$_x$ for free,
periodic, and twisted periodic longitudinal boundary conditions.  The term
``twisted'' means that the longitudinal ends of the strip are identified with
reversed orientation.  These strip graphs can be embedded on surfaces with the
following topologies:\footnote{\footnotesize{For the triangular lattice with
cylindrical BC's, $W$ and ${\cal B}$ were calculated in \cite{baxter}. Other
calculations of $P$, $W$, and ${\cal B}$ have been performed for strips having
BC's of type (i) \cite{strip}-\cite{w2d}, (ii) \cite{w2d}, (iii)
\cite{pg}-\cite{nec}, (iv) \cite{pg,pm}.}}  : (i) (FBC$_y$,FBC$_x$): strip;
(ii) (PBC$_y$,FBC$_x$): cylindrical; (iii) (FBC$_y$,PBC$_x$): cylindrical
(denoted cyclic here); (iv) (FBC$_y$,TPBC$_x$): M\"obius; (v)
(PBC$_y$,PBC$_x$): torus; and (vi) (PBC$_y$,TPBC$_x$): Klein bottle.\footnote{
\footnotesize{These BC's can all be implemented in a manner that is uniform in
the length $L_x$; the case (vii) (TPBC$_y$,TPBC$_x$) with the topology of the
projective plane requires different identifications as $L_x$ varies and will
not be considered here. For connections between topology and graph properties,
see e.g. \cite{sk,wb}.}}  Here we present and analyze chromatic polynomials for
the strip graph of the square lattice with $L_y=3$ (i.e. cross sections forming
triangles) and boundary conditions of type (v) and (vi): torus and Klein
bottle.  We recall that unlike graphs of type (i)-(v), the Klein bottle 
graph (vi) cannot be embedded without self-intersection in ${\mathbb R}^3$. 
For $L_x=m \ge 2$ where they are well defined, the $L_y=3$
torus and Klein bottle graphs have $n=L_xL_y$ vertices, $e=2n$ edges, the same
girth $g$ (length of minimum closed circuit) and number $k_g$ of circuits of
length $g$, and the respective chromatic numbers $\chi=3$ and $\chi=4$.

We label a particular type of strip graph as $G_s$ and the specific graph of
length $L_x=m$ repeated subgraph units, e.g. columns of squares in the case of
the square strip, as $(G_s)_m$.
If one thinks of the graph as embedded on a rectangular 
strip of paper, with its upper and lower sides glued together and its 
longitudinal ends glued with direct or reversed orientation, then $L_x$ is the 
length of this strip of paper in subgraph units.  Writing 
\beq
P((G_s)_m,q) = \sum_{j=0}^{n-1}(-1)^j h_{n-j}q^{n-j}
\label{p}
\eeq
and using the results that \cite{m,rtrev} 
$h_{n-j}={e \choose j}$ for $0 \le j < g-1$ (whence $h_n=1$ and $h_{n-1}=e$) 
and $h_{n-(g-1)}={e \choose g-1}-k_g$, it follows that for $m$ greater than the
above-mentioned minimal value, these $h_j$'s are the same for the torus and 
Klein bottle of each type $G_s$.  For a given $G_s$, as $m$ increases, the 
$h_{n-j}$'s for the torus and Klein bottle graphs become equal for larger $j$. 

A generic form for chromatic polynomials for recursively defined families of 
graphs, of which strip graphs $G_s$ are special cases, is
\beq
P((G_s)_m,q) =  \sum_{j=1}^{N_\lambda} c_j(q)(\lambda_j(q))^m
\label{pgsum}
\eeq 
where $c_j(q)$ and the $N_\lambda$ terms $\lambda_j(q)$ depend on the type of 
strip graph $G_s$ but are independent of $m$.  

For an $L_y=3$, $L_x=m$ strip with (PBC$_y$,FBC$_x$) one has \cite{w2d} 
$P(sq(L_y=3)_m,PBC_y,FBC_x,q) = q(q-1)(q-2)(q^3-6q^2+14q-13)^{m-1}$,
whence
\beq
W(sq(L_y=3),PBC_y,FBC_x,q) = (q^3-6q^2+14q-13)^{1/3}
\label{wsqly3pbc}
\eeq
with ${\cal B}=\emptyset$. 

In order to calculate $P$, one may use recursive methods based on iterative use
of deletion-contraction theorems \cite{bds,bm,strip} or a 
coloring compatibility matrix method described in 
\cite{matmeth,bgen}.  For the $L_y=3$ torus ($t$) graphs, one finds
\beq
P(sq(L_y=3)_m,PBC_y,PBC_x,q) = \sum_{j=1}^8 c_{t,j} (\lambda_{t,j})^m
\label{ptorus}
\eeq
where
\beq
\lambda_{t,1}=-1 \ , \qquad c_{t,1} = q^3-6q^2+8q-1 \ ,
\label{lamtor1}
\eeq
\beq
\lambda_{t,2}=q^3-6q^2+14q-13 \ , \qquad c_{t,2} = 1 \ ,
\label{lamtor2}
\eeq
\beq
\lambda_{t,3}=q-1 \ , \qquad c_{t,3}=\frac{(q-1)(q-2)}{2} \ ,
\label{lamtor3}
\eeq
\beq
\lambda_{t,4}=q-4 \ , \qquad c_{t,4}=(q-1)(q-2) \ ,
\label{lamtor4}
\eeq
\beq
\lambda_{t,5}=q-2 \ , \qquad c_{t,5}=q(q-3) \ , 
\label{lamtor5}
\eeq
\beq
\lambda_{t,6}=q-5 \ , \qquad c_{t,6}=\frac{q(q-3)}{2} \ ,
\label{lamtor6}
\eeq
\beq
\lambda_{t,7}=-(q^2-7q+13) \ , \qquad c_{t,7}=q-1 \ , 
\label{lamtor7}
\eeq
\beq
\lambda_{t,8}=-(q-2)^2 \ , \qquad c_{t,8}=2(q-1) \ . 
\label{lamtor8}
\eeq

For the $L_y=3$ Klein $(K)$ bottle graphs one finds 
\beq
P(sq(L_y=3)_m,PBC_y,TPBC_x,q) = \sum_{j=1}^5 c_{K,j} (\lambda_{K,j})^m
\label{pklein}
\eeq
where
\beq
\lambda_{K,1}=\lambda_{t,1}=-1 \ , \qquad c_{K,1} = -(q-1) \ ,
\label{lamk1}
\eeq
\beq
\lambda_{K,2}=\lambda_{t,2}=q^3-6q^2+14q-13 \ , \qquad c_{K,2} = c_{t,2}=1 \ ,
\label{lamk2}
\eeq
\beq
\lambda_{K,3}=\lambda_{t,3}=q-1 \ , 
\qquad c_{K,3}=-c_{t,3}=-\frac{(q-1)(q-2)}{2} \ ,
\label{lamk3}
\eeq
\beq
\lambda_{K,4}=\lambda_{t,6}=q-5 \ , \qquad c_{K,4}=c_{t,6}=\frac{q(q-3)}{2} \ ,
\label{lamk4}
\eeq
\beq
\lambda_{K,5}=\lambda_{t,7}=-(q^2-7q+13) \ , \qquad c_{K,5}=c_{t,7}=q-1 \ . 
\label{lamk5}
\eeq
The terms $\lambda_{t,j}$, $j=4,5,8$ do not enter in eq. (\ref{pklein}).  We
contrast this with earlier findings.  For a given strip, $N_\lambda$ was found
to be larger for (FBC$_y$,PBC$_x$) than (FBC$_y$,FBC$_x$) \cite{strip,pg,wcy}.
For the $L_y=3$ square lattice strip case, $N_\lambda=2$ for (FBC$_y$,FBC$_x$)
but $N_\lambda=1$ for (PBC$_y$,FBC$_x$), because of the special feature that
the cross sections were complete graphs, $K_p$ with $p=3$ and hence the
intersection theorem led to a factorized, monomial form for $P$.  It was found
\cite{wcy,pm} that for a given type of lattice strip, $N_\lambda$ is the same
for the (FBC$_y$,PBC$_x$) = cyclic and (FBC$_y$,TPBC$_x$) = M\"obius
topologies, although the $c_j$'s were, in general, different. The present
results show that reversal of orientation in the identification of opposite
ends of a strip can lead to a change in $N_\lambda$.\footnote{\footnotesize{A
different sort of change in $P$, accompanied by a change in ${\cal B}$, can be
obtained if one considers a homogeneous recursive family and the same family
with a finite inhomogeneous subgraph inserted, e.g., the ``rope ladder'' graphs
of \cite{read91} or two such subgraphs forming ends, viz., the $J(\prod H)I$
strip graphs in \cite{strip}.}}

Let $C=\sum_{j=1}^{N_\lambda} c_j$. We find $C=P(K_3,q)=q(q-1)(q-2)$ for the
$L_y=3$ torus graphs and $C=0$ for the $L_y=3$ Klein bottle graphs.  The zero
results from the special constraints introduced by the boundary conditions and
is analogous to the fact that $C=0$ for the $L_y=2$ M\"obius square strip
\cite{bds} and its homeomorphic expansions \cite{pg}. However, not all M\"obius
strip graphs have $C=0$; for example, for the $L_y=3$ M\"obius strips of the
square and kagom\'e lattices, $C=q(q-1)$ and $C=q$, respectively \cite{pm}.

\begin{figure}
\vspace{-4cm}
\centering
\leavevmode
\epsfxsize=4.0in
\begin{center}
\leavevmode
\epsffile{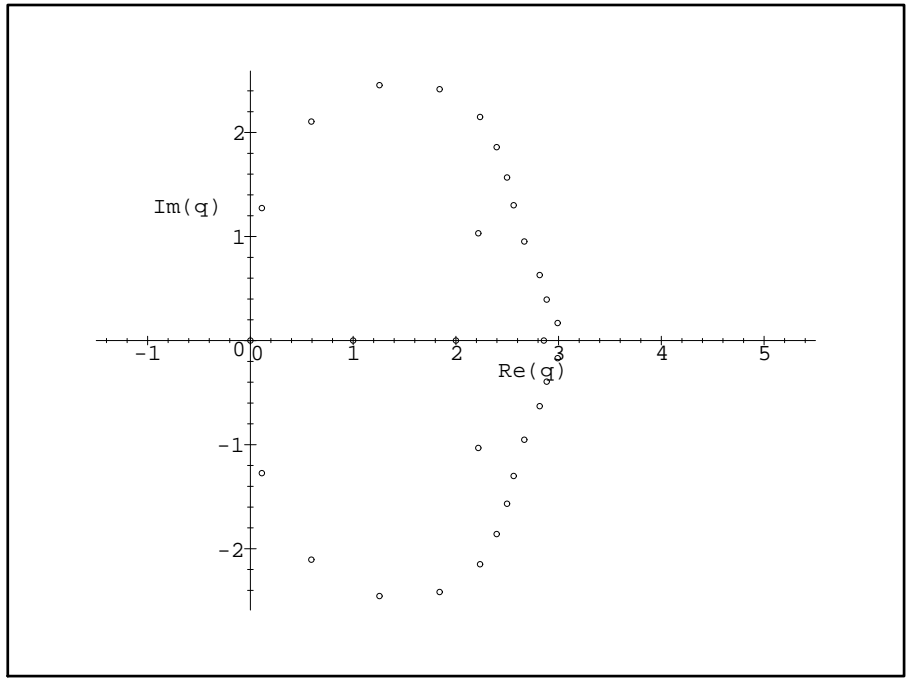}
\end{center}
\vspace{-2cm}
\caption{\footnotesize{Chromatic zeros for the $L_y=3$, $L_x=m=10$ torus 
graph.}}
\label{toruszeros}
\end{figure}

Chromatic zeros for the $L_y=3$, $m=10$ torus graph are shown in Fig.
\ref{toruszeros}; with this value of $m$, the chromatic zeros for the Klein
bottle graph are quite similar.  The locus ${\cal B}$ and the $W$ functions are
the same for the torus and Klein bottle graph families.  We find $q_c=3$,
which, interestingly, is the same value as for the infinite 2D square lattice.
The locus ${\cal B}$ has support for $Re(q) \ge 0$ and separates the $q$ plane
into three regions.  The outermost one, region $R_1$, extends to infinite $|q|$
and includes the intervals $q \ge 3$ and $q \le 0$ on the real $q$ axis.
Region $R_2$ includes the real interval $2 \le q \le 3$ and extends upward and
downward to the complex conjugate triple points on ${\cal B}$ at $q_t$ and
$q_t^*$, where $q_t \simeq 2.5 + 1.4i$.  Region $R_3$ is the innermost one and
includes the real interval $0 \le q \le 2$.  The boundary between $R_2$ and
$R_3$ curves to the right as one increases $|Im(q)|$, extending from $q=2$
upward to $q_t$ and downward to $q_t^*$.  As is evident in
Fig. \ref{toruszeros}, the density of chromatic zeros on the $R_1-R_3$ 
boundary near $q=0$ and on the $R_3-R_2$ boundary is somewhat smaller than on
the right-hand part of the $R_2-R_1$ boundary. 

In region $R_1$, $\lambda_{t,2}=\lambda_{K,2}$ is the dominant $\lambda_j$, so 
\beq
W = (q^3-6q^2+14q-13)^{1/3} \ , \quad q \in R_1 \ . 
\label{wr1}
\eeq
The fact that this is the same as $W$ for the (PBC$_y$,FBC$_x$) case, eq. 
(\ref{wsqly3pbc}), is a general result.  The importance of the PBC$_y$ is 
evident from the fact that for the same width of three squares, the strip
with (FBC$_y$,FBC$_x$) yields a different $W$ \cite{strip}. 

In region $R_2$ $\lambda_{t,6}=\lambda_{K,4}$ is dominant, so
\beq
|W| = |q-5|^{1/3} \ , \quad q \in R_2
\label{wr2}
\eeq
(in regions other than $R_1$, only $|W|$ can be determined unambiguously 
\cite{w}). In region $R_3$, $\lambda_{t,7}=\lambda_{K,5}$ is dominant, so 
\beq
|W|=|q^2-7q+13|^{1/3} \ , \quad q \in R_3 \ . 
\label{wr3}
\eeq 
The outer boundary separating $R_1$ from the inner two regions is oblate,
extending out to a maximum of about $|Im(q)| \simeq 2.5$ for $Re(q) \simeq 1.5$
(and passing through $q=0$ and 3).  All the three points, $q=0,2,3$, where
${\cal B}$ crosses the real $q$ axis, it does so vertically.  The present
results are in accord with the inference \cite{strip,wcy} that for a recursive
graph with regular lattice structure, a necessary and sufficient condition for
${\cal B}$ to separate the $q$ plane into two or more regions is that it
contains a global circuit, i.e. a path along a lattice direction whose length
goes to infinity as $n \to \infty$; here this is equivalent to PBC$_x$.  The
fact that ${\cal B}$ is the same for these torus and Klein families means that
none of $\lambda_{t,j}$, $j=4,5,8$ is a dominant term. 

Our calculations of the zero-temperature Potts antiferromagnet partition
functions (chromatic polynomials) and exponential of the entropy, $W$, for
$L_y=3$ strips of the square lattice with periodic transverse and periodic and
twisted periodic longitudinal boundary conditions (torus and Klein bottle
graphs) thus elucidates the role that these boundary conditions and the
associated topologies play; the torus and Klein bottle graphs have 
interestingly different chromatic polynomials, with different $N_\lambda$, 
but the $W$ functions and hence the boundaries ${\cal B}$ are the same.

\vspace{4mm}

The research of R. S. was supported in part by the U. S. NSF grant 
PHY-97-22101.

\vfill
\eject

\begin{thebibliography}{99}

\bibitem{birk}{Birkhoff, G. D. 1912 Ann. of Math. {\bf 14}, 42.}

\bibitem{rrev}{Read, R. C. 1968 J. Combin. Theory {\bf 4}, 52.}

\bibitem{rtrev}{Read, R. C. and Tutte, W. T. 1988 ``Chromatic Polynomials'',
in {\it Selected Topics in Graph Theory, 3}, (Academic Press, New York), 
p. 15.} 

\bibitem{bbook}{Biggs, N. L. 1993 {\it Algebraic Graph Theory} (Cambridge 
Univ. Press, Cambridge).}

\bibitem{potts}{Potts, R. B. 1952 Proc. Camb. Phil. Soc. {\bf 48}, 106.}

\bibitem{wurev}{Wu, F. Y. 1982 Rev. Mod. Phys. {\bf 54}, 235.} 

\bibitem{w}{Shrock, R. and Tsai, S.-H. 1997 Phys. Rev. {\bf E55}, 5165.}

\bibitem{bkw}{Beraha, S., Kahane, J., and Weiss, N. 1980 J. Combin. Theory B
{\bf 28}, 52.}

\bibitem{bds}{Biggs, N. L., Damerell, R. M. and Sands, D. A. 1972
J. Combin. Theory B {\bf 12}, 123.}

\bibitem{bm}{Biggs, N. L. and Meredith, G. H. 1976 J. Combin. Theory B 
{\bf 20}, 5.}

\bibitem{b}{Biggs, N. L. 1977 Bull. London Math. Soc. {\bf 9}, 54.} 

\bibitem{matmeth}{Biggs, N. L. LSE report LSE-CDAM-99-03 (May 1999), to 
appear.} 

\bibitem{wcy}{Shrock, R. and Tsai, S.-H. 1999 Phys. Rev. E, in press; 
Stony Brook report ITP-SB-99-12 (April, 1999).} 

\bibitem{pm}{Shrock, R., Stony Brook report ITP-SB-99-23 (June, 1999).} 

\bibitem{readcarib88}{Read, R. C. 1988 in Proc. 5th Caribbean Conf. on
Combin. and Computing.} 

\bibitem{read91}{Read, R. C. and Royle, G. F. 1991 in {\it Graph Theory,
Combinatorics, and Applications} (Wiley, NY), vol. 2, p. 1009.}

\bibitem{baxter}{Baxter, R. J. 1987 J. Phys. A {\bf 20}, 5241.}

\bibitem{strip}{Ro\v{c}ek, M., Shrock, R., and Tsai, S.-H. 1998 Physica
{\bf A252}, 505; {\it ibid.} {\bf A259}, 367.}

\bibitem{hs}{Shrock, R. and Tsai, S.-H. 1998 Physica {\bf A259}, 315.}

\bibitem{w2d}{Shrock, R. and Tsai S.-H. 1998 Phys. Rev. {\bf E58}, 4332,
cond-mat/9808057.}

\bibitem{pg}{Shrock, R. and Tsai, S.-H. 1999 J. Phys. A Lett. {\bf 32} L195.}

\bibitem{nec}{Shrock, R. and Tsai, S.-H. 1999 J. Phys. A, in press
(cond-mat/9905431).} 

\bibitem{sk}{Saaty, T. L. and Kainen, P. C. 1977 {\it The Four-Color Problem} 
(McGraw-Hill, New York), p. 45.}

\bibitem{wb}{White, A. T. and Beineke, L. W. 1978 in Beineke, L. W. and Wilson,
R. J. {\it Selected Topics in Graph Theory} (Academic, New York) p. 15.} 

\bibitem{m}{Meredith, G. H. 1972 J. Combin. Theory B {\bf 13}, 14.}

\bibitem{bgen}{Biggs, N. L. LSE report LSE-CDAM-99-05 (June 1999).} 

\end{thebibliography}
\end{document}